\documentstyle[12pt,epsfig]{article}
\begin{document}
\vspace{1.cm}
\begin{center}
\    \par
\    \par
\    \par
\     \par

        {\Large { \bf{  Study of collective matter flow in central C-Ne and
           C-Cu collisions  at energy
                of 3.7 GeV per nucleon }}}
\end{center}
\   \par
\   \par
\   \par
\   \par
\    \par
\par
{\large { \bf{L.Chkhaidze, T.Djobava, L.Kharkhelauri}}}\par
\   \par
\   \par
\    \par
{\it {High Energy Physics Institute, Tbilisi State University}},\par
{\it {University St 9, 380086 Tbilisi, Republic of Georgia}}\par
{\it {Fax: (99532) 30-98-52;}}  \par
{\it {E-mail: djobava@sun20.hepi.edu.ge or}}\par
{\it {\hspace{1.5cm}                     ida@sun20.hepi.edu.ge}} \par
\pagebreak
\begin{center}

                     \bf{ ABSTRACT }
\end{center}
\ \par
\par
    The transverse momentum technique is used to analyse  charged-particle
exclusive data in the central C-Ne and C-Cu interactions at energy of
3.7 Gev per nucleon.
The clear evidence of in-plane and out-of-plane (squeeze-out)
flow effects for
protons and $\pi^{-}$ mesons have been obtained.
In C-Ne interactions the $\pi^{-}$ mesons in-plane flow is in the same direction
to the protons, while in C-Cu collisions pions show antiflow behaviour.
From the transverse momentum and azimuthal distributions of
 protons and $\pi^{-}$ mesons with respect
to the reaction plane the  flow $F$ (the measure of the amount
of collective transverse momentum transfer in the reaction plane) and
the parameter $a_{2}$ (the measure of the strength of the anisotropic emission)
have been extracted. The flow effects increase with the mass of the particle
and atomic number of target $A_{T}$.
The comparison of our in-plane flow results with flow data of various
projectile/target configurations had been done by a scaled flow
$F_{S}$=$F/(A_{P}^{1/3}+A_{T}^{1/3})$. $F_{S}$ demonstrates a common scaling
behaviour among flow values from different systems.
\par
\     \par
\     \par
PACS numbers: 25.70.-z; 25.75.Ld
\pagebreak
\   \par
\   \par
\   \par
\par
One of the main goal of relativistic heavy-ion collisions
experiments is to study nuclear matter under extreme condistions of high
density and temperatures, i.e. to learn more about the nuclear equation
of state (EOS).
An increasing number of observables which are accessible through
heavy-ion collisions has been found to be sensitive to the EOS.
In order to study the EOS, collective effects, such as the bounce-off of cold
spectator matter in the reaction plane [1] --- the directed transverse flow
and the squeeze-out of hot and compressed participant matter perpendicular
to the reaction plane [2] -- the elliptic flow are frequently used.
    According to theoretical
    calculations the studies of flow can provide information  on the
    collision dynamics as well as on
    a possible phase transition to soft quark matter.
\par
    Collective flow is the consequence of the pressure
    buildup in the high density zone through the short range repulsion
    between nucleons, i.e. through compressional energy. This effect
    leads to characteristic, azimuthally asymmetric sidewards emission
    of the reaction products.
     While the transverse
    flow in the reaction plane is influenced by the cold matter
    deflected by the overlap region of the colliding nuclei, the squeeze
    out is caused by the hot and compressed matter from the interaction
    region which preferentially escapes in the direction perpendicular
    to the reaction plane unhindered by the presence of the projectile
    and target spectators.
\par
The efforts to determine the EOS and the more general aspect of producing
high-energy densities over extended regions have led to a series of
experiments to study relativistic nucleus-nucleus collisions
at BEVALAC (Berkeley), GSI-SIS (Darmstadt), JINR (Dubna),
AGS (Brookhaven National Laboratory) and SPS (CERN).
\par
Using
the transverse momentum technique developed by P.Danielewicz
and G.Odyniec [3], nuclear collective flow has already been observed
for protons, light nuclei, pions and $\Lambda$ - hyperons emitted in
nucleus-nucleus
collisions at energies 0.4$\div$1.8 GeV/nucleon of BEVALAC, GSI-SIS [4-12],
 at 11$\div$14 GeV/nucleon of AGS [13,14] and at 158 GeV/nucleon
of CERN [15]. The discovery
of collective sidewards flow in Au+Au at the AGS was a major
highlight at 1995 [14].
\par
In this article we present experimental results obtained from the in
 and out-plane
transverse momentum analysis for protons and $\pi^{-}$ mesons
in central C-Ne and C-Cu
interactions at energy E=3.7 GeV per nucleon
with the SKM-GIBS set-up of JINR. The signature for collective flow
had been obtained. It shows the persistence of collective flow phenomena
all the way up to AGS energies. The observed results
obtained by streamer chamber technique  allow to extend the experimental data available
from BEVALAC, GSI-SIS  and AGS. These results provide quantitative information on
the transverse and out-of-plane  (squeeze-out) elliptic flows and their dependence
on beam energy and projectile/target mass.
\par
 SKM-GIBS consists of a 2 m streamer chamber, placed in a magnetic field
of 0.8 T, and a triggering system. The streamer chamber was exposed to
beam of C nuclei accelerated in the synchrophasotron
up to energy of 3.7 GeV/nucleon. The thickness of the solid target
(of the form of thin disc) -- Cu was 0.2 g/cm$^{2}$.
 Neon gas filling of the chamber
also served as a nuclear target.
 The triggering system  allowed  the  selection  of
"inelastic"  and "central" collisions.
\par
   The inelastic trigger was selecting all inelastic  interactions of
incident nuclei on a target.
\par
   The central trigger was selecting events with no charged projectile
spectator fragments (with  $P/Z>3$ GeV/c ) within a cone of  half angle
$\Theta_{ch}$ = 2.4$^{0}$ or  2.9$^{0}$   (the trigger efficiency was
99$\%$ for events  with  a single charged particle in the cone). The biases
and correction procedures were discussed in detail in ref. [16,17]. The
ratio  $\sigma_{cent}$/$\sigma_{inel}$  (that characterizes the
centrality of selected events) is  - (9$\pm$1)$\%$ for C-Ne
and (21$\pm$3)$\%$ - for C-Cu.
In Table 1 the number of events are presented. Average
measurement errors of the momentum and production angle determination
for protons are $<\Delta P/P>$= (8$\div$10)$\%$, $\Delta$$\Theta$ =1$^{0}$$\div$2$^{0}$
and for pions are ~~~$<\Delta P/P>$= 5$\%$, $\Delta$$\Theta$ =0.5$^{0}$.
\par
The data have been analysed event by event using the transverse momentum
technique of P.Danielewicz and G.Odyniec [3].
Using this method, nuclear collective flow
for protons has been observed
in central C-Ne and C-Cu interactions at
a momentum of P=4.5 GeV/c/N (E=3.7 GeV/nucl) with the SKM-200 set-up of JINR
and presented in our previous paper[18]. The results presented there for protons in C-Cu interactions
are obtained on a statistics two times larger than in [18] and both results
for C-Ne and
C-Cu collisions are represented in terms of the normalized rapidity $y/y_{p}$
($y_{p}$-- projectile rapidity, $y_{p}$=2.28)
in the laboratory system unlike of [18].
P.Danielewicz and G.Odyniec  have proposed
an exclusive way to analyse the momentum contained in directed sidewards
emission and present the data in terms of the mean transverse momentum
per nucleon in the reaction plane $<P_{x}(Y)>$ as a function of the rapidity.
The vector
$\overrightarrow{Q_{j}}=\sum\limits_{i\not=j}\limits^{n} \omega_{i}
\overrightarrow{P_{{\perp}i}}$ was used for the reaction plane
(the reaction plane is the plane containing
$\overrightarrow{Q_{j}}$
and the beam axis)
determination
of each event,
where $P_{{\perp}i}$ is the transverse momentum  of particle $i$, and $n$ is
the number of particles in the event.
 Pions are not included.
The weight $\omega_{i}$ is the function
$\omega_{i}$= $y_{i}$ - $<y>$ as in [9], where  $<y>$ is the
average rapidity, calculated for each event over all the participant protons,
i.e. protons which are not fragments of the projectile ($P/Z>3$ GeV/c,
$\Theta < 4^{0}$) and
target ($P<0.2$ GeV/c).
The average multiplicities of analysed  protons $<N_{p}>$
 are listed in Table 1.
 The transverse momentum of each
particle  in the estimated reaction plane is calculated as
$P_{xj}\hspace{0.01cm}^{\prime} = \{ \overrightarrow{{Q_{j}}}\cdot
\overrightarrow{P_{{\perp}j}} /
\vert\overrightarrow{{{Q_{j}}}} $
\par
 The average transverse momentum
$<P_{x}\hspace{0.01cm}^{\prime}(Y)>$
%$<P_{x}(Y)>$
is obtained by averaging over all
events in the corresponding intervals of rapidity.
\par
For the event by event analysis it is necessary to perform an
identification of $\pi^{+}$ mesons, the admixture of which
amongst the charged positive particles is about (25$\div$27)$\%$ . The
identification has been carried out on the statistical basis using
 the two-dimentional ( $P_{\parallel}$, $P_{\perp}$ ) distribution.
It had been assumed, that  $\pi^{-}$  and $\pi^{+}$ mesons
hit a given cell of the plane
( $P_{\parallel}$, $P_{\perp}$ ) with equal probability.
 The difference in multiplicity of $\pi^{+}$  and $\pi^{-}$
in each event was required to be no more than 2. After
this procedure the admixture of $\pi^{+}$ is not exceeding
(5$\div$7)$\%$.
 The temperature of the identified protons agrees with our
previous result [19], obtained by the subtraction method of spectra.
\par
It is known [4], that the estimated reaction plane
differs from the true  one,
due to the finite number of particles in each event.
The component $ P_{x}$ in the true reaction plane is systematically larger
then the component $P_{x}\hspace{0.01cm}^{\prime}$
in the estimated plane, hence
$<P_{x}>=<P_{x}\hspace{0.01cm}^{\prime}>/<cos\varphi> $,
where $\varphi$ is the angle between the estimated and true planes.
The  correction factor
$K$=1 $/$ $< cos\varphi >$ is subject to a large uncertainty,
especially for low multiplicity.
For the definition of $< cos\varphi>$ according to [3], we divided
randomly each event into two equal sub-events.
The values of $K$,
averaged over all the multiplicities, are: $K=1.27\pm0.08$ ---
for C-Ne,  $K=1.31\pm0.04$ --- for C-Cu.
\par
Fig.1  show the dependence of $<P_{x}>$
on the normalized rapidity $y/y_{p}$ in the laboratory system
for protons and pions
in C-Ne (Fig.1.a) and C-Cu (Fig.1.b)
collisions. For protons the data points are
already corrected  (multiplied by $K$)
for the deviation from the true reaction plane.
The data exhibit the typical $S$-shape behaviour
which demonstrates the collective transverse momentum transfer between the
forward and backward hemispheres.
\par
From the mean transverse momentum distributions we can extract  an
observable -- the transverse flow
$F=<P_{x}>/d(y/y_{p})$, i.e. the slope
of the momentum distribution at midrapidity. It is a measure of the amount
of collective transverse
momentum transfer in the reaction. Technically $F$ is obtained by fitting
the central part of the dependence of
$<P_{x}>$ on $y/y_{p}$
by the first order polynomial function,
 this coefficient is the flow $F$.
The fit was done for $y/y_{p}$ between 0.01 $\div$ 0.90.
The straight lines in Fig.1 show the results of this fit.
The values of $F$ are listed in Table 1.
We have analysed the influence of the admixture
of ambiguously identified $\pi^{+}$ mesons  on the results.
 The error in flow $F$ includes the statistical and systematical
 errors.
One can see from the Table 1, that with the increase of the atomic number of
the target A$_{T}$, the value of $F$ increases. A similar
tendency had been observed at lower energies [4-7,10].
\par
It is of great  interest to compare the flow values for a wide range of data.
A way of comparing the energy dependence of flow values for different
projectile/target mass combinations was suggested by A.Lang et al [20] and
first used by J.Chance in [10]. To allow for
different projectile/target ($A_{P}$,$A_{T}$)  mass systems, they divided
the flow values by ($A_{P}^{1/3}+A_{T}^{1/3}$) and called
$F_{S}=F/(A_{P}^{1/3}+A_{T}^{1/3})$ the scaled flow.
\par
Fig.2 shows a plot of $F_{S}$ versus energy per
nucleon of the projectile. We have included our data, the data from the EOS
[10,21], E-895, E-877 [21],
FOPI [12] experiments along with the values derived from the
Plastic Ball [7,11] and
the Streamer Chamber [4,9] experiments for a variety of energies and mass
combinations.
The values of flow $F$ for E-895, E-877 are taken from Fig.5 of [21].
Then these values are recalculated in terms of $F_{S}$.
For the EOS and the Plastic Ball data all the isotopes of
$Z$=1 and 2 are included, except for [11] where the data is for $Z$=1.
 The Streamer Chamber data [4,9] normally include
all protons, whether free or bound in clusters as in our case.
In Fig.2 the scaled flow values
$F_{S}$ follow, within the uncertainties, a common trend with an initial
steep rise and then an indication of a gradual decrease.
It is worth to mention,
that the data obtained by streamer chamber technique (including our results)
are somewhat (slightly) larger than ones obtained by the electronic experiments.
This is caused may be by the small mixture of bounded protons (deutons, $^{3}H$,
$^{4}He$).
\par
In our previous
paper [18]  the  collective flow for the protons in C-Ne and C-Cu
interactions have been compared with the predictions of the Quark Gluon
String Model (QGSM). The QGSM reprodused the experimental results , but
underestimated the values of flow  $F$.
\par
 In view of the large number of pions created at our energies and reactions
and the strong coupling between the nucleon and pion, it is interesting to
know if pions also have a collective flow behaviour and if yes, how the pion flow
is related to the nucleon flow.
\par
For this purpose the reaction plane have been obtained by (over) the protons
and the transverse momentum of each $\pi^{-}$ meson have been  projected onto
this reaction plane.
Fig.1 show the dependence of $<P_{x}>$
on the normalized rapidity $y/y_{p}$ in the laboratory system
for $\pi^{-}$ mesons
in C-Ne and C-Cu collisions.
The data exhibit the typical $S$-shape behaviour as for the protons.
The values of flow $F$ for $\pi^{-}$ mesons
are: for C-Ne collisions $F=29 \pm 5$ MeV;
for C-Cu --- $F=-47 \pm 6$ MeV.
The straight lines in Fig.1 show the results of this fit.
The fit was done in the following intervals of $y/y_{p}$: 0.04 $\div$ 0.7
for C-Ne ;  -0.06 $\div$ 0.6  for C-Cu .
The absolute value of $F$ increases
with the atomic number of target $A_{T}$, which indicates on the rise of
collective flow effect. The similar tendency have been obtained in [8]
for $\pi^{-}$ and $\pi^{+}$ mesons in Ne-Naf, Ne-Nb and Ne-Pb interactions
at 800 MeV/nucleon energy.
\par
 One can see from the Fig.1, that for C-Ne collisions the $ \overrightarrow
{P_{x}}$ of the pions is directed in the same direction as protons i.e.
flow of protons and pions are correlated, while for C-Cu interactions the
$ \overrightarrow{P_{x}}$ of $\pi^{-}$ mesons is directed oppositely to that of
the protons (antiflow).
\par
At AGS energy of 11 GeV/nucleon [22] have been obtained that the flow
of $\pi^{+}$ mesons is in the direction
opposite to the protons, similar to observations
 in semi-central
Pb-Pb collisions at energy 158 GeV/nucleon in WA98 collaboration at SPS CERN
[15]. The magnitude of the directed flow in [15] is found to be
significantly smaller than observed at AGS energies.  Thus it seems that
the flow effects for the pions decreases with the increasing the energy.
Theoretical calculations using
the Isospin Quantum Molecular Dynamics (IQMD) model
have predicted [23] the existence of pion antiflow at projectile- and
target rapidities
for Au-Au collisions at GSI-SIS  energies -- 1 GeV/nucleon. On the other hand
within the framework of the relativistic transport model (ART 1.0) [24] for
heavy-ion collisions (Au-Au) at AGS energies, pions are found to have a weak
flow behaviour.
\par
The origin of the particular
shape of the $\overrightarrow{P_{x}}$ spectra for pions had been studied
in [23-26].
The investigation revealed, that
the origin of
the  in-plane transverse momentum of pions is
the pion scattering process (multiple $\pi N $ scattering) [23]
and the pion absorption [25,26].  However in [24] had been found, that
the pions show a weak flow behaviour in central collisions  due to the
flow of baryon resonances from which they are produced.
\par
The anticorrelation of nucleons and pions in [23] was explained as due
to multiple $\pi ~N$ scattering. However in [24,26] it had been shown, that
the anticorrelation is a manifestion of the nuclear shadowing effect
of the target- and projectile-spectator  through both pion rescattering and
reabsorptions.  In our opinion, our results indicate, that the flow behaviour
of $\pi^{-}$ mesons in light system -- C-Ne is due to the flow of
$\Delta$ resonances, whereas the antiflow behaviour in C-Cu collisions is the
result of the nuclear shadowing effect.
\par
The preferential emission of particles in the direction perpendicular to the
reaction plane (i.e. "squeeze-out")
is particularly interesting since it is only way where nuclear
matter might escape without being rescattered by spectator remnants of
the projectile and target and is expected to provide direct information
on the hot and dense participant region formed in high energy
nucleus-nucleus interactions. This phenomenon, predicted by hydrodynamical
calculations [2], has been clearly identified in the experiments [27]
by the observation of an enhanced out-of-plane emission of
protons, mesons and charged fragments. For beam energies of 1 $\div$ 11
GeV/nucleon the elliptic flow results from a strong competition between
the early "squeeze-out" and the late stage "in-plane flow" [28]. The magnitude
and the sign of elliptic flow depend on two factors: a) the pressure built up
in the compression stage compared to the energy density and b) the passage
time of the projectile and target spectators.
\par
In order to extend these
investigations,
the azimuthal $\phi$  ( $cos\phi=P_{x}/P_{t}$) distributions
of the pions and protons with respect to the reaction plane
 have been studied.
The angle $\phi$ is the
relative azimuthal angle between the true
reaction plane and the emitted particle.
 To select particles emitted from the participant
zone, the analysis was restricted only to the mid-rapidity region by applying
a cut around the center of mass rapidity.
Fig.3 show respective
distributions for protons and $\pi^{-}$ mesons in C-Ne (Fig.3.a)
and C-Cu (Fig.3.b) collisions.
For visual representation the data of C-Cu are shifted up.
For $\pi^{-}$ mesons the analysis was performed from 0 to 180$^{0}$ due to the
lower (smaller) statistics then for the protons. The azimuthal angular distributions
for the protons and pions show a maxima at $\phi$=90$^{0}$  and
$270^{0}$ with respect to the event plane.
This maxima is associated with preferential particle emission perpendicular to
the reaction plane (squeeze-out, or elliptic flow).  Thus a clear signature of an
out-of plane signal is
evidenced.
\par
To treat the data in a quantitative way the azimuthal distributions had been
fitted by Fourier series:
$dN/d\phi=a_{0}(1+a_{1}cos\phi+a_{2}cos2\phi)$
\par
The anisotropy factor $a_{2}$ is negative for out-of-plane enhancement
(squeeze-out) and is the measure of the strength of the anisotropic
emission.
The values of the coefficients $a_{2}$  extracted from the
azimuthal distributions of protons and $\pi^{-}$ mesons are presented
in Table 2.
 The fitted curves are superimposed
on the experinetal distributions (Fig.3).
\par
The values $a_{2}$ are used to quantify the ratio $R$ of the number of particles
emitted perpendicular to the number of particles emitted in the reaction plane,
which represents the magnitude of the out-of-plane emission signal:
$R=(1-a_{2})/(1+a_{2})$. A ratio $R$ larger than unity implies a preferred
out-of-plane emission. The values of $R$ are listed in Table 2 respectively.
One can see from Table 2, that the $a_{2}$ and $R$ increases both for protons
and $\pi^{-}$ mesons with:
narrowing the cut applied around the center of mass rapidity
(protons in C-Cu interactions); increasing the transverse momentum
and the atomic number of target $A_{T}$. The squeeze-out effect
is more pronounced for protons than for $\pi^{-}$ mesons.
Our results concerning rapidity, mass and transverse momentun
dependence of the azimuthal anisotropy are consistent with analysis from the
Plastic Ball, FOPI, Kaos, TAPS [27,29] collaborations and
are confirmed by IQMD calculations [30].
\par
In the experiments (E-895, E-877, EOS) [31] at AGS and SPS (CERN) (NA49)
energies the elliptic flow is typically studied at midrapidity and quantified
in terms of the second Fourier coefficient $v_{2}\approx<cos2\phi>$. The Fourier
coefficient $v_{2}$ is related to $a_{2}$ via the equation $v_{2}=a_{2}/2$.
We have estimated $v_{2}$ for C-Ne and C-Cu. The dependence of the elliptic flow
excitation function (for protons) on energy E$_{lab}$ is displayed in Fig.4.
 Recent calculations have made specific predictions for the
beam energy dependence of elliptic flow for Au-Au collisions at 1 $\div$ 11
GeV/nucleon [28]. They indicate a transition from negative to positive elliptic
flow at a beam energy E$_{tr}$, which has a marked sensitivity to the
stiffness of the EOS. In addition, they suggest that a phase transition to the
Quark-Gluon Plasma (QGP) should give a characteristic signature in the elliptic
flow excitation function due to the significant softening of the EOS. One can
see from Fig.4, that the excitation function $v_{2}$ clearly shows an
evolution from negative to posotive elliptic flow within the region 2 $\leq
E_{beam} \leq $ 8 GeV/nucleon and point to an apparent transition energy
E$_{tr} \sim $ 4 GeV/nucleon.
\par
In summary, in this paper we have reported experimental results,
presented
in terms of the mean transverse momentum per nucleon projected onto the
 reaction plane $<P_{x}>$
as a function of the  normalized rapidity $y/y_{p}$ in laboratory system.
We have determined the flow  $F$, defined as the slope
at midrapidity.
The $F$ increases
with the mass of the particle and atomic number of target $A_{T}$.
In C-Ne interactions the $\pi^{-}$ mesons flow is in the same direction
to the protons, while in C-Cu collisions pions show antiflow behaviour.
The comparison of our in-plane flow results with flow data of various
projectile/target configurations have been done by a scaled flow
$F_{S}$=$F/(A_{P}^{1/3}+A_{T}^{1/3})$. $F_{S}$ demonstrates a common scaling
behaviour among flow values from different systems.
From the azimuthal distributions of protons and $\pi^{-}$ mesons with respect
to the reaction plane at mid-rapidity region
a clear signature of an out-of-plane flow (squeeze-out)
have been obtained. The azimuthal distributions have been parametrized
by second order Fourier series and the parameter $a_{2}$ of the
anisotropy term $a_{2}cos2\phi$ have been extracted.
The $R=(1-a_{2})/(1+a_{2})$ ratio have been calculated. The squeeze-out effect
increases with the transverse momentum,
atomic number of target $A_{T}$ and also by narrowing the
applied cut around the center mass rapidity. It is more pronounced
for protons than for $\pi^{-}$ mesons.
\    \par
\   \par
\   \par
\   \par
%\newpage
ACKNOWLEDGEMENTS
\par
\   \par
\   \par
 We would like to thank M.Anikina, A.Golokhvastov, S.Khorozov and J.Lukstins
for  fruitful collaboration during the obtaining of the data.
We are very grateful to
Z. Menteshashvili for his continuous support
during the
preparation of the article.
\pagebreak
\par

\newpage
\begin{center}
\bf{FIGURE CAPTIONS}
\end{center}
{\bf{Fig.1}}
The dependence of $< P_{x} >$ on the normalized rapidity
 $y/y_{p}$ in the laboratory system
 in C-Ne (a) and C-Cu (b) collisions.
$\circ$ -- for protons,
$\bigtriangleup$ -- for $\pi^{-}$ mesons.
The
lines are the result of the approximation of experimental data
 of protons and $\pi^{-}$ mesons by
first order polynom in the interval of
 $0.01\leq y/y_{p} \leq0.90$ for protons (C-Ne, C-Cu)
and $0.04\leq y/y_{p} \leq0.70$ (C-Ne), $-0.06\leq y/y_{p} \leq0.6$ (C-Cu) for
$\pi^{-}$ mesons
. The curves for
visual presentation
- result of approximation data by 4-th order polynoms.
\   \par
{\bf{Fig.2}}
Scaled flow values versus beam energy per nucleon for different
projectile/target systems.
\mbox{\put(3.,0.){\framebox(6.,6.)[cc]{}}}~~~~ -- Nb-Nb Plastic Ball
 $\bigtriangleup$ -- Au-Au Plastic Ball,
 $\circ$ -- Ni-Ni FOPI,
$\bullet$ -- Ni-Cu EOS, + -- Au-Au EOS,
$\oplus$ -- Ni-Au EOS,
$\star$ -- Ar-Pb Streamer Chamber, the value at E=1.08 AGeV represents Ar-KCl
Streamer Chamber,
 $\diamond$ -- C-Ne, C-Cu  our result, $\ast$ -- Au-Au E-895, the value at
E=10 AGeV represents Au-Au from E-877.
% $\times$ -- Pb-Pb NA49.
 To improve the distinction between data
points at the same beam energy, some of the beam energy values have been shifted.
\   \par
{\bf{Fig.3}}
The azimuthal distributions with respect to the reaction plane of midrapidity
protons dN/d$\phi$ (a) and $\pi^{-}$ mesons (b).
$\circ$ -- for C-Ne ($-1\leq y_{cm}\leq1$),
$\bigtriangleup$ -- for C-Cu ($-1\leq y_{cm}\leq1$) interactions.
The curves --- result of approximation by
$dN/d\phi=a_{0}(1+a_{1}cos\phi+a_{2}cos2\phi)$.
\   \par
{\bf{Fig.4}}
The dependence of the elliptic flow excitation function v$_{2}$ on energy
E$_{lab}$/A (GeV).
$\star$ -- FOPI, $\circ$ -- MINIBALL, $\bullet$ -- EOS,
\mbox{\put(3.,0.){\framebox(6.,6.)[cc]{}}}~~~~ -- E-895,
$\times$ -- E-877, $\oplus$ -- NA49, $\bigtriangleup$ -- C-Ne, C-Cu our results.
\newpage
%\   \par
%\   \par
%\    \par
\begin{center}
\bf{TABLE CAPTION}
\end{center}
\   \par
{\bf{Table 1}}. The number of experimental
events,  the average  multiplicity of participant
protons $<N_{p}>$,
 the correction factor $K$ and
 the flow  $F$ for protons and $\pi^{-}$ mesons.\\
\   \par
{\bf{Table 2}}. The values of the parameter $a_{2}$ and the ratio $R$
for protons and $\pi^{-}$ mesons extracted
from the azimuthal distributions  fitted by
$dN/d\phi=a_{0}(1+a_{1}cos\phi+a_{2}cos2\phi)$.

\newpage
Table 1. The number of experimental
events,  the average  multiplicity of participant
protons $<N_{p}>$,
 the correction factor $K$ and
 the flow  $F$ for protons and $\pi^{-}$ mesons.
\   \par
\   \par
\   \par
\   \par
\   \par
\   \par
\begin{tabular}{|l|c|c|}    \hline
&     &   \\
\hspace{0.5cm} &\hspace{1.cm} C-Ne \hspace{1.3cm} &\hspace{1.3cm} C-Cu\hspace{1.6cm}        \\
&     &   \\
\hline
  Number of exper. events  &      723      &      667      \\
\hline
    $<N_{p}>$          &  12.4 $\pm$ 0.5   &  19.5$\pm$ 0.6   \\
\hline
 $K$=1/$< cos\varphi>$  &  1.27 $\pm$ 0.08 &  1.31 $\pm$ 0.04   \\
\hline
  $F$  for protons (MeV/c)   &  134  $\pm$12  & 198$\pm$13     \\
\hline
  $F$ for $\pi^{-}$ mesons (MeV/c) &   29 $\pm$5    & -47$\pm$6     \\
\hline
\end{tabular}
\newpage
Table 2. The values of the parameter $a_{2}$ and the ratio $R$
for protons and $\pi^{-}$ mesons extracted
from the azimuthal distributions  fitted by
$dN/d\phi=a_{0}(1+a_{1}cos\phi+a_{2}cos2\phi)$.
\   \par
\   \par
\   \par
\   \par
\   \par
\   \par
\begin{tabular}{|c|c|c|c|c|}    \hline
&  &  &  &  \\
$A_{p}-A_{T}$ &Particle  &Applied Cut   & $a_{2}$& $R$ \\
&  &  &  &  \\
\hline
&Protons&$-1\leq y_{cm}\leq1$&-0.049$\pm$0.014& 1.10$\pm$0.03 \\
\cline{3-5}
&  & $-1\leq y_{cm}\leq1$; $P_{T}\geq 0.3$ GeV/c&-0.074$\pm$0.014&1.16$\pm$0.04 \\
\cline{2-5}
C-Ne& $\pi^{-}$ mesons& $-1\leq y_{cm}\leq1$&-0.035$\pm$0.013&1.07$\pm$0.04 \\
\cline{3-5}
&   & $-1\leq y_{cm}\leq1$; $P_{T}\geq 0.1$ GeV/c & -0.050$\pm$0.014& 1.09$\pm$0.03\\
\hline
&  & $-1\leq y_{cm}\leq1$& -0.065$\pm$0.014&1.14$\pm$0.04 \\
\cline{3-5}
&Protons  & $-1\leq y_{cm}\leq1$; $P_{T}\geq 0.3$& -0.081$\pm$0.014&1.18$\pm$0.05 \\
\cline{3-5}
C-Cu&  & $-0.6\leq y_{cm}\leq0.6$& -0.077$\pm$0.017& 1.17$\pm$0.04\\
\cline{3-5}
& & $-0.6\leq y_{cm}\leq0.6$; $P_{T}\geq 0.3$& -0.088$\pm$0.020&1.19$\pm$0.06 \\
\cline{2-5}
& $\pi^{-}$ mesons& $-1\leq y_{cm}\leq1$&-0.041$\pm$0.013&1.08$\pm$0.03 \\
\cline{3-5}
&   & $-1\leq y_{cm}\leq1$; $P_{T}\geq 0.1$ GeV/c & -0.056$\pm$0.015&1.12$\pm$0.04\\
\hline
\end{tabular}
%\end{document}
\pagebreak
\begin{figure}
\begin{center}
\epsfig{file=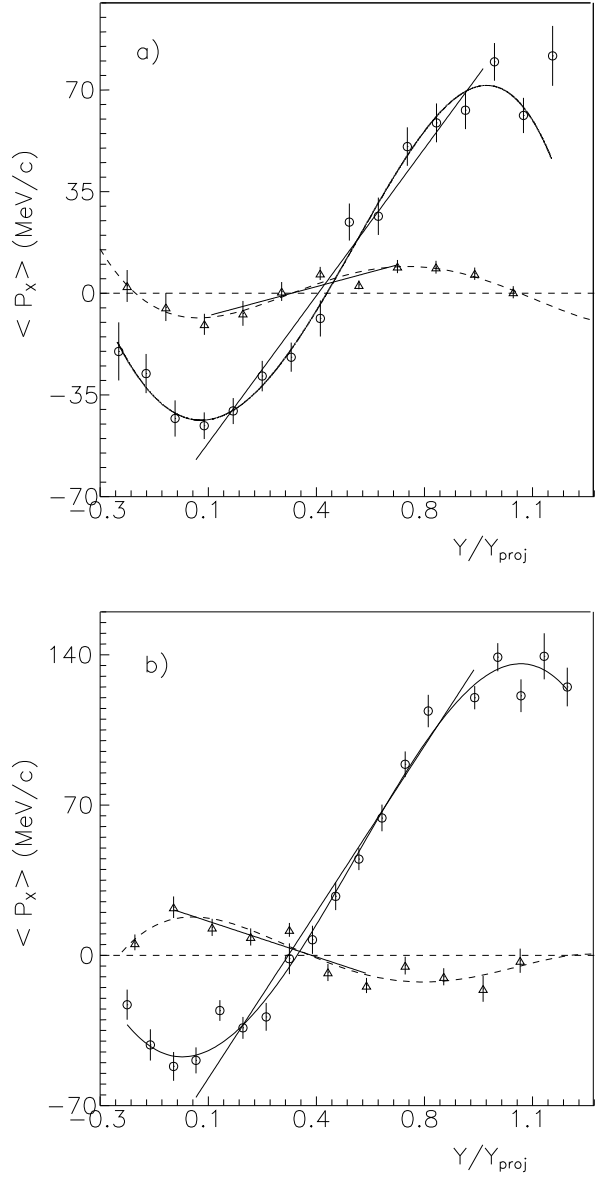,bbllx=0pt,bblly=0pt,bburx=594pt,bbury=842pt,
width=18cm,angle=0}
\end{center}
\vspace{-7.9cm}
\hspace{0.cm}
\begin{minipage}{16.cm}
\caption
{The dependence of $< P_{x} >$ on the normalized rapidity
 $y/y_{p}$ in the laboratory system
 in C-Ne (a) and C-Cu (b) collisions.
$\circ$ -- for protons,
$\bigtriangleup$ -- for $\pi^{-}$ mesons.
The
lines are the result of the approximation of experimental data
 of protons and $\pi^{-}$ mesons by
first order polynom in the interval of
 $0.01\leq y/y_{p} \leq0.90$ for protons (C-Ne, C-Cu)
and $0.04\leq y/y_{p} \leq0.70$ (C-Ne), $-0.06\leq y/y_{p} \leq0.6$ (C-Cu) for
$\pi^{-}$ mesons
. The curves for
visual presentation
- result of approximation data by 4-th order polynoms.}
\end{minipage}
\end{figure}
%-----------------------------------------------------
\pagebreak
\begin{figure}
\begin{center}
\epsfig{file=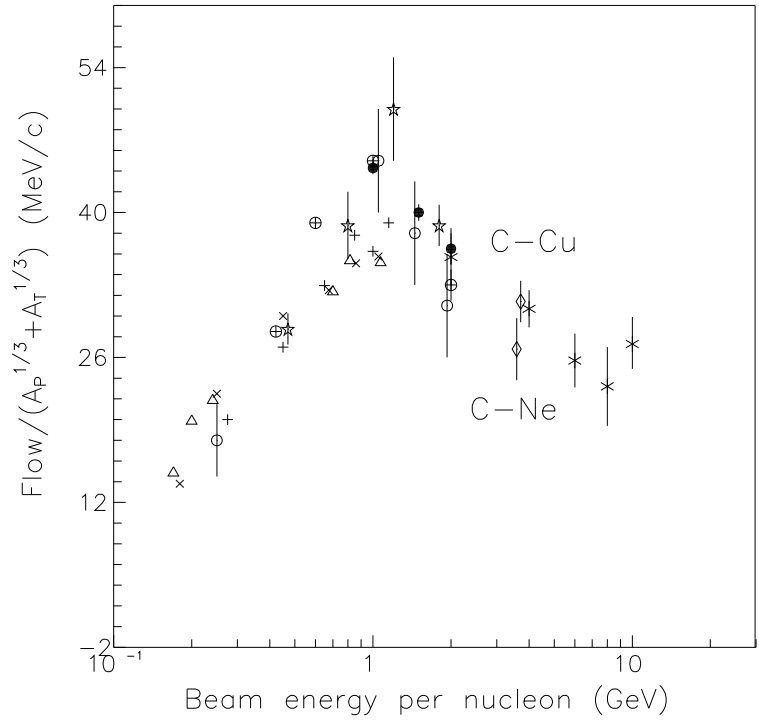,bbllx=0pt,bblly=0pt,bburx=594pt,bbury=842pt,
width=18.0cm,angle=0}
\end{center}
\vspace{-10.cm}
\hspace{3.cm}
\begin{minipage}{10.0cm}
\caption
{Scaled flow values versus beam energy per nucleon for different
projectile/target systems.
%\mbox{\put(3.,0.){\framebox(6.,6.)[cc]{}}}~~~~ 
%${\framebox(5,5){}}$
$\times$ -- Nb-Nb Plastic Ball
 $\bigtriangleup$ -- Au-Au Plastic Ball,
 $\circ$ -- Ni-Ni FOPI,
$\bullet$ -- Ni-Cu EOS, + -- Au-Au EOS,
$\oplus$ -- Ni-Au EOS,
$\star$ -- Ar-Pb Streamer Chamber, the value at E=1.08 AGeV represents Ar-KCl
Streamer Chamber,
 $\diamond$ -- C-Ne, C-Cu  our result, $\ast$ -- Au-Au E-895, the value at
E=10 AGeV represents Au-Au from E-877.
%$\times$ -- Pb-Pb NA49.
 To improve the distinction between data
points at the same beam energy, some of the beam energy values have been 
shifted.}
\end{minipage}
\end{figure}
%------------------------------------------------------
%\pagebreak
\pagebreak
\begin{figure}
\begin{center}
\epsfig{file=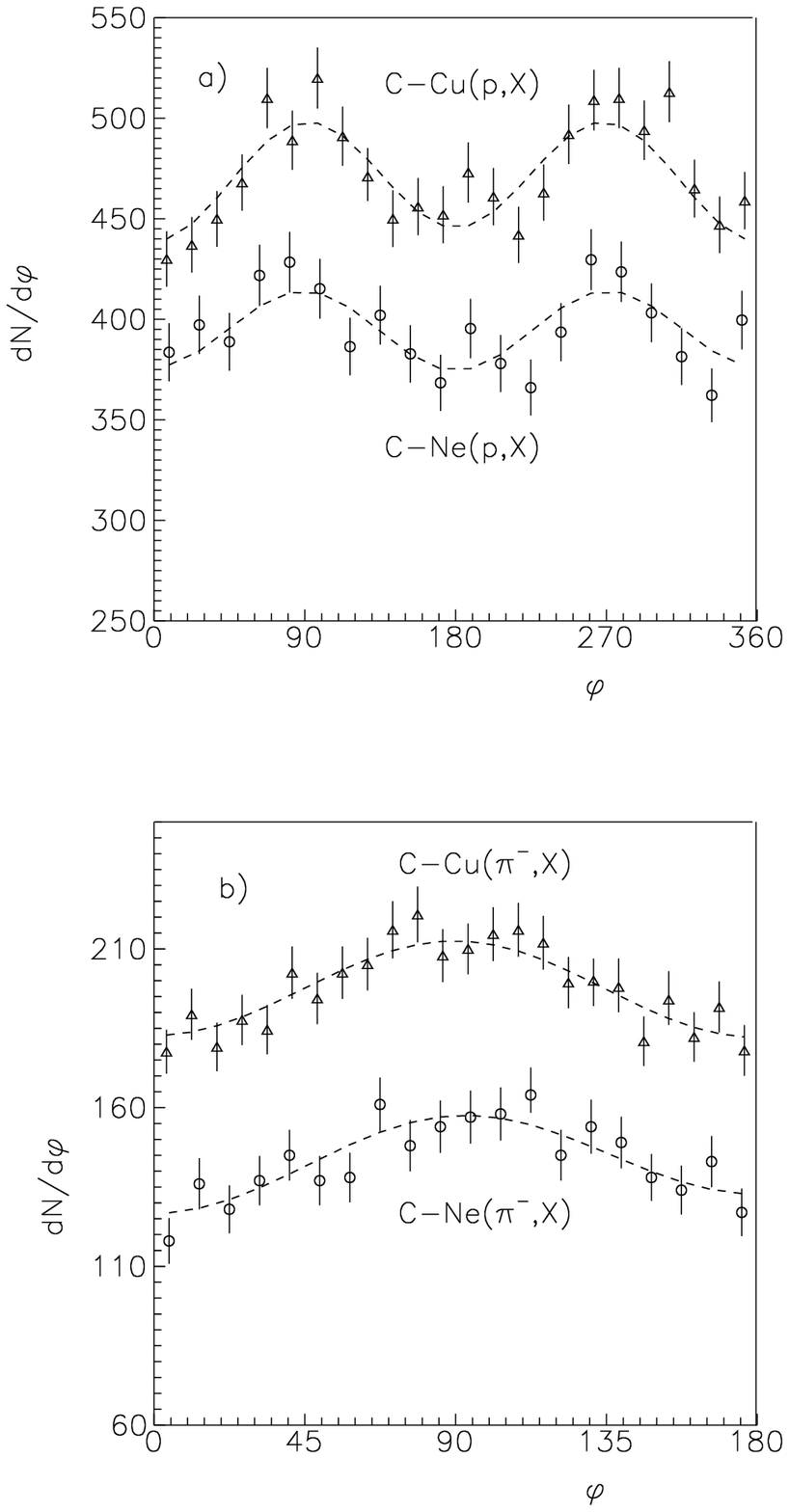,bbllx=0pt,bblly=0pt,bburx=594pt,bbury=842pt,
width=18cm,angle=0}
\end{center}
\vspace{-6.4cm}
\hspace{0.cm}
\begin{minipage}{16.0cm}
\caption
{The azimuthal distributions with respect to the reaction plane of midrapidity
protons dN/d$\phi$ (a) and $\pi^{-}$ mesons (b).
$\circ$ -- for C-Ne ($-1\leq y_{cm}\leq1$),
$\bigtriangleup$ -- for C-Cu ($-1\leq y_{cm}\leq1$) interactions.
The curves --- result of approximation
$dN/d\phi=a_{0}(1+a_{1}cos\phi+a_{2}cos2\phi)$.}
\end{minipage}
\end{figure}
%------------------------------------------------------------
%\end{document}
\pagebreak
\begin{figure}
\begin{center}
\epsfig{file=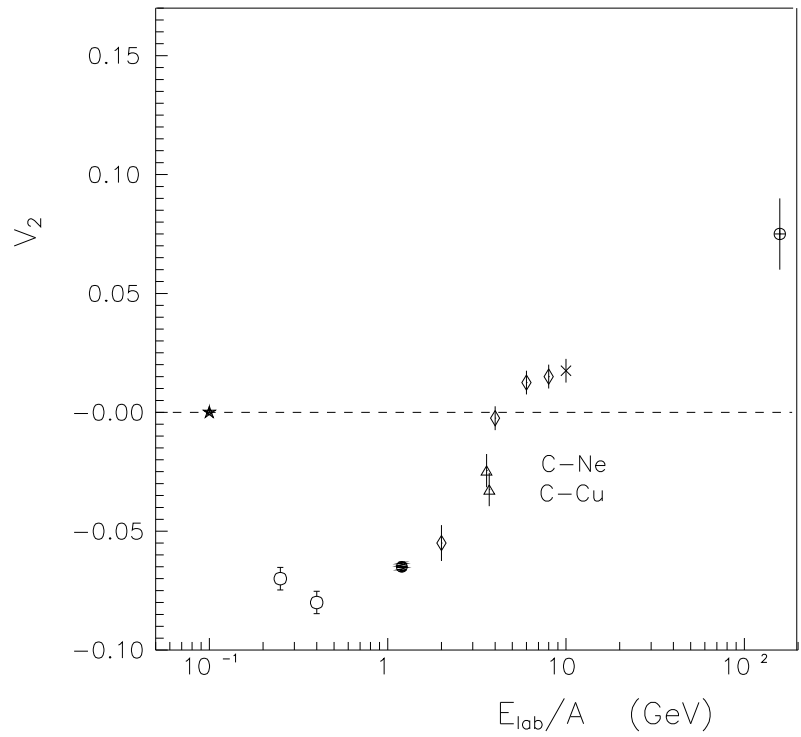,bbllx=0pt,bblly=0pt,bburx=594pt,bbury=842pt,
width=18cm,angle=0}
\end{center}
\vspace{-6.4cm}
\hspace{0.cm}
\begin{minipage}{16.0cm}
\caption
{The dependence of the Elliptic flow excitation function v$_{2}$ on energy
E$_{lab}$/A (GeV).
$\star$ -- FOPI, $\circ$ -- MINIBALL, $\bullet$ -- EOS,
%\mbox{\put(3.,0.){\framebox(6.,6.)[cc]{}}}~~~~ 
%$\framebox(5,5){}}$
$\diamond$ -- E-895,
$\times$ -- E-877, $\oplus$ -- NA49,
$\bigtriangleup$ -- C-Ne, C-Cu our results.}
\end{minipage}
\end{figure}
\end{document}